\documentclass[aps,pra,twocolumn,superscriptaddress]{revtex4-2}

\usepackage[english]{babel}
\usepackage{amsmath}
\usepackage{amsthm}
\usepackage{amssymb}
\usepackage{mathrsfs}
\usepackage{booktabs}
\usepackage{tikz}
\usepackage{color}
\usepackage{hyperref}
\usepackage[ruled,vlined]{algorithm2e}
\usepackage{cases}
\usepackage{dblfloatfix}
\usepackage{multirow}
\usepackage{bbm}
\usepackage{enumitem}
\usepackage{caption}
\usepackage[font=small,labelfont=bf,justification=raggedright]{caption}
\hypersetup{
  colorlinks   = true,  
  urlcolor     = blue,  
  linkcolor    = blue,  
  citecolor    = red    
}
\usepackage{mathtools}
\usepackage{natbib}
\usepackage{pgfplots}
\usepackage{siunitx}
\usepackage{subfig}

\usepackage{url}
\newcommand{\ket}[1]{ | #1 \rangle }
\newcommand{\bra}[1]{ \langle #1 | }

\newcommand{\di}{{\rm d}}

\mathtoolsset{showonlyrefs}

\usepackage{scalerel,stackengine}
\stackMath
\newcommand\reallywidehat[1]{%
\savestack{\tmpbox}{\stretchto{%
  \scaleto{%
    \scalerel*[\wi\di thof{\ensuremath{#1}}]{\kern-.6pt\bigwedge\kern-.6pt}%
    {\rule[-\textheight/2]{1ex}{\textheight}}
  }{\textheight}%
}{0.5ex}}%
\stackon[1pt]{#1}{\tmpbox}%
}
\usepackage{comment}
\usepackage{float}

\SetArgSty{textnormal}

\makeatletter

\newenvironment{widetext2}{%
  \par\ignorespaces
  \setbox\widetext@top\vbox{%
   \vskip15\p@
   \hb@xt@\hsize{%
    \leaders\hrule\hfil
    \vrule\@height6\p@
   }%
   \vskip6\p@
  }%
  \setbox\widetext@bot\hb@xt@\hsize{%
    \vrule\@depth6\p@
    \leaders\hrule\hfil
  }%
  \onecolumngrid
  \let\set@footnotewidth\set@footnotewidth@ii
}{%
  \par
  \twocolumngrid\global\@ignoretrue
  \@endpetrue
}%

\makeatother

\begin{document}
\title{$\Lambda$-Enhanced Gray Molasses Cooling of $^{85}$Rb Atoms in Tweezers Using the D$_2$ Line}

\author{D.A. \surname{Janse van Rensburg}}
\thanks{These authors contributed equally to this work}

\author{R.C. \surname{Venderbosch}}
\thanks{These authors contributed equally to this work}
\author{Y. \surname{van der Werf}}
\author{J.J. \surname{del Pozo Mellado}}
\author{M.L. \surname{Venderbosch}}
\author{R.S. \surname{Lous}}
\author{E.J.D. \surname{Vredenbregt}}
\author{S.J.J.M.F. \surname{Kokkelmans}}

\affiliation{Coherence and Quantum Technology group, Department of Applied Physics and Science Education, Eindhoven University of Technology, P.O. Box 513, 5600 MB Eindhoven, The Netherlands}
\affiliation{Center for Quantum Materials and Technologies Eindhoven (QT/e), Eindhoven University of Technology, P.O. Box 513, 5600 MB Eindhoven, The Netherlands}

\date{\today}

\begin{abstract}
We demonstrate the implementation of $\Lambda$-enhanced gray molasses cooling on the D$_2$ line of $^{85}$Rb atoms in an optical tweezer array. This technique yields lower atomic temperatures of $\SI{4.0(2)}{\micro\kelvin}$ compared to red-detuned polarization gradient cooling, and consequently extends the $T_2^*$ coherence time of the hyperfine clock qubit by a factor of 1.5. The method is alignment-free and can be readily implemented on laser beams used for magneto-optical trapping, as it only requires frequency and phase modulation control. Our experimental observations are corroborated by a numerical model based on a semi-classical force approach extended to a four-level system, including two hyperfine states of the upper manifold that are $\SI{120}{\mega\hertz}$ apart.
\end{abstract}

\maketitle

\section{Introduction}
Neutral atoms trapped in optical tweezers have shown promise as a platform for quantum simulation and quantum information experiments \cite{10.1116/5.0036562,RevModPhys.82.2313}, with atoms of the alkali metal group commonly used for this purpose \cite{bluvstein_logical_2024,66s8-jj18, manetsch_tweezer_2025}. For those platforms, efficient sub-Doppler cooling is a requirement for achieving long coherence times, since the far red-detuned optical tweezers impose a differential light shift on the atoms. This results in dephasing that is temperature dependent for non-magic wavelength traps \cite{PhysRevA.72.023406, GRIMM200095,magictrapping}.

A versatile approach to reduce the temperature of single alkali atoms in optical tweezers -- and thereby extend their coherence times -- is $\Lambda$-enhanced gray molasses cooling ($\Lambda$-GMC) \cite{graham_multi-qubit_2022,evered_high-fidelity_2023}. This scheme combines blue-detuned gray molasses cooling on a type-II transition ($F'\leq F$) with velocity-selective coherent population trapping to minimize photon scattering events for the cold atomic population \cite{PhysRevA.87.063411}. $\Lambda$-GMC is typically executed on the D$_1$ line due to the favorable three-level structure \cite{Salomon_2013,rio_fernandes_sub-doppler_2012,PhysRevA.94.033408,PhysRevA.93.023421}.

$\Lambda$-GMC can also be implemented efficiently on the D$_2$ line, if the hyperfine splitting is much larger than the transition linewidth $\Gamma$. The technique has been successfully implemented on $^{87}$Rb \cite{D2LambdaCooling}, $^{133}$Cs \cite{PhysRevA.98.033419} and $^{40}$K \cite{bruce_sub-doppler_2017} alkali metal isotopes on the D$_2$ line, reaching temperatures of several $\si{\micro\kelvin}$. In contrast, the small $\sim6\,\Gamma$ hyperfine splitting of $^{23}$Na limits the achievable temperature reached by $\Lambda$-GMC on the D$_2$ line to \SI{56}{\micro\kelvin} \cite{Shi_2018}, whereas \SI{7}{\micro\kelvin} \cite{PhysRevA.93.023421} has been achieved using the D$_1$ line. Still, the use of the D$_2$ line is convenient for most alkali atoms as the required laser is typically already used for a MOT, although it introduces the complication that blue-detuned light for the type-II $F\rightarrow F'=F$ transition is red-detuned from the type-I $F\rightarrow F'=F+1$ transition.

In this article, we demonstrate $\Lambda$-GMC on the D$_2$ line of $^{85}$Rb, and explore the cooling efficiency in a four-level structure. This transition has a natural linewidth of $\Gamma/2\pi = \SI{6.06669(18)}{\mega\hertz}$ \cite{Steck}, and features an intermediate hyperfine splitting of $19.9\,\Gamma$ between the type-I and type-II transitions. Furthermore, we demonstrate improved coherence times on the hyperfine clock qubit due to optimized $\Lambda$-GMC. To complement the experimental results, we have generalized the theoretical description of the $\Lambda$-GMC to a four-level system relevant to the D$_2$ line of alkali metal atoms.

\section{$\Lambda$-GMC Demonstration}
\subsection{Method}
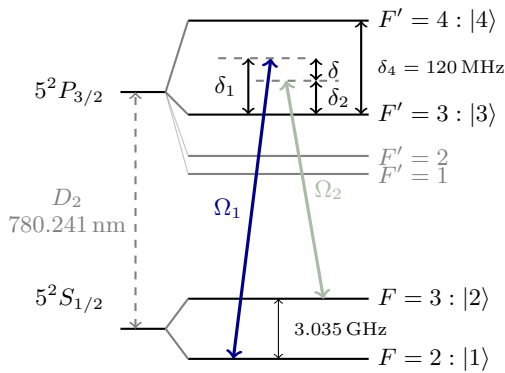
\begin{figure}
    \centering
    \begin{tikzpicture}[scale=1.0, font=\small]

\tikzset{
  level/.style={thick},
  connect/.style={gray, thick},
  detuning/.style={gray, dashed, thick, -{Latex[length=2mm]}},
}

\draw[level] (0,-0.4) -- (2.4,-0.4) node[right]{$F=2:\ket{1}$};
\draw[level] (0,0.4) -- (2.4,0.4) node[right]{$F=3:\ket{2}$};
\draw[level] (-0.9,0.0) -- (-0.3, 0.0);
\draw[-,gray,thick] (-0.3,0.0) -- (0,0.4);
\draw[-,gray,thick] (-0.3,0.0) -- (0,-0.4);
\draw[<->] (1.2,-0.4) -- (1.2,0.4) node[midway,right]{\begin{scriptsize}
    3.035\,GHz
\end{scriptsize}};

\draw[level, opacity = 0.5] (0,2.06) -- (2.4,2.06) node[right]{$F'=1$};
\draw[level, opacity = 0.5] (0,2.3) -- (2.4,2.3) node[right]{$F'=2$};
\draw[level] (0,2.85) -- (2.4,2.85) node[right]{$F'=3:\ket{3}$};
\draw[level] (0,4.1) -- (2.4,4.1) node[right]{$F'=4:\ket{4}$};
\draw[level] (-0.9,3.15) -- (-0.3, 3.15);
\draw[-,gray,thick] (-0.3,3.15) -- (0,2.85);
\draw[-,gray,thick] (-0.3,3.15) -- (0,4.1);
\draw[-,gray, opacity = 0.5] (-0.3,3.15) -- (0,2.3);
\draw[-,gray, opacity = 0.5] (-0.3,3.15) -- (0,2.06);
\draw[<->, thick] (2.3,2.85) -- (2.3,4.1) node[midway,right]{\begin{scriptsize}
    $\delta_4=120$\,MHz
\end{scriptsize}};

\node[left] at (-1,0.4) {$5^2S_{1/2}$};
\node[left] at (-1,3.15) {$5^2P_{3/2}$};

\draw[<->,gray,thick,dashed] (-0.7,0.0) -- (-0.7,3.15)
  node[midway,left,align=center,gray]{$D_2$\\780.241\,nm};

\draw[<->, color={rgb,255:red,0; green,0; blue,137}, very thick] (0.6,-0.4) -- (1.1,3.6)node[midway, left]{$\Omega_1$};
\draw[<->, color={rgb,255:red,170; green,187; blue,166}, very thick] (1.8,0.4) -- (1.3,3.3) node[midway, right]{$\Omega_2$};

\draw[<->,thick] (0.8,2.85) -- (0.8,3.6)
  node[midway,left=1pt]{$\delta_1$};
\draw[<->,thick] (1.7,2.85) -- (1.7,3.3)
  node[midway,right=1pt]{$\delta_2$};
  \draw[<->,thick] (1.7,3.3) -- (1.7,3.6)
  node[midway,right=1pt]{$\delta$};

\draw[-,gray,thick, dashed] (0.4,3.6) -- (1.6,3.6);
\draw[-,gray,thick, dashed] (0.9,3.3) -- (2.0,3.3);

\end{tikzpicture}
    \caption{\textbf{D$_2$ line of $^{85}$Rb, focusing on the four levels \{$\ket{1},\ket{2}, \ket{3}, \ket{4}$\} involved in $\Lambda$-GMC.} The parameters $\{\delta_1,\Omega_1\}$ represent the repumper detuning and Rabi frequency (blue, $F=2 \rightarrow F'=3)$, the repumper light is an EOM sideband on the cooling laser. The parameters $\{\delta_2,\Omega_2\}$ are the cooling laser carrier detuning and Rabi frequency (green, $F=3\rightarrow F'=3$), $\delta = \delta_1-\delta_2$ is the Raman detuning, and $\delta_4$ the hyperfine splitting.}
    \label{fig:energydiagramexp}
\end{figure}

We implement $\Lambda$-GMC in our neutral atom setup \cite{vanrensburg2025fidelityrelationsarrayneutral,deonThesis}, which traps single atoms in a 10$\times$10 array of far red-detuned, $\SI{813}{\nano\meter}$, optical tweezers. The core laser in the setup is \SI{780}{\nano\meter}, which addresses the D$_2$ line of $^{85}$Rb, as shown in the energy-level diagram in Fig. \ref{fig:energydiagramexp}. This laser is used for creating a magneto-optical trap, imaging single atoms, optical pumping into the 5$^2$S$_{1/2}\ket{F=2,m_F=0}$ state for qubit operations, qubit state projection via a hyperfine-selective blowout pulse, and $\Lambda$-GMC. We use an electro-optic modulator (EOM) to produce frequency sidebands for the $5^2$S$_{1/2}\ket{F=2}$ $\leftrightarrow$ $5^2$P$_{3/2}\ket{F'=3}$ repumper transition during the MOT stage, imaging of the single atoms, and $\Lambda$-GMC.

The beam delivery system for $\Lambda$-GMC does not require hardware upgrades on the setup and is directly implemented on existing 3D MOT beams. During $\Lambda$-GMC, a repumper detuning $\delta_1$ is tuned to the Raman resonance condition and can be scanned over a wide range. We operate an acousto-optic modulator (AOM) to scan the carrier detuning $\delta_2$ over a range from $+6\,\Gamma$ to $+14.3\,\Gamma$ from the $5^2$S$_{1/2}\ket{F=3}$ $\leftrightarrow$ $5^2$P$_{3/2}\ket{F'=3}$ transition maintaining consistent laser power levels. The MOT is created at $-3\,\Gamma$ from the free-space $5^2$S$_{1/2}\ket{F=3}$ $\leftrightarrow$ $5^2$P$_{3/2}\ket{F'=4}$ transition.

We investigate the cooling behavior using the recapture probability after a \SI{40}{\micro\second} release time to study the relative heating and cooling of the atoms when addressing multiple transitions of the D$_2$ line. The initial temperature of the atoms before a \SI{20}{\milli\second} $\Lambda$-GMC pulse is \SI{9.7(4)}{\micro\kelvin} determined by the release and recapture method \cite{PhysRevA.78.033425}.

\subsection{Experimental Results}
\label{sec:ExpResults}
As a first scan, we vary the repump detuning for fixed carrier detuning over the hyperfine manifold as shown in Fig.~\ref{fig:lgmexp}a. We use a carrier detuning of $\delta_2=6\,\Gamma$ with a Rabi frequency of $\Omega_2=1.63\,\Gamma$, and a repumper-to-carrier ratio of $(\Omega_1/\Omega_2)^2=0.1$. The typical recapture probability of the atoms at the initial temperature before cooling is indicated by the dashed black line.

We observe heating when the light is red-detuned, and slight cooling when it is blue-detuned with respect to the S$_{1/2}\ket{F=2}\leftrightarrow$ P$_{3/2}\ket{F' = 1}$ and S$_{1/2}\ket{F=2}\leftrightarrow$ P$_{3/2}\ket{F' = 2}$ transitions. We attribute the heating to the fact that most scattering events occur on these \lq repumper' transitions, which are type-II transitions where sub-Doppler cooling arises when the light is blue-detuned. This is similar to the condition that both the cooling and repumper light must be blue-detuned of their respective transitions to achieve the coldest temperatures in a blue-detuned MOT of alkali metal atoms \cite{PhysRevLett.120.083201}. Near resonance with the S$_{1/2}\ket{F=2}\leftrightarrow$ P$_{3/2}\ket{F' = 3}$ transition $(\delta\approx-\delta_2)$ we observe pronounced cooling, which we attribute to gray molasses cooling. Finally, near the Raman resonance condition ($\delta=0$) we observe the characteristic narrow linewidth cooling and heating feature of $\Lambda$-GMC.

 We perform similar measurements for various carrier and Raman detunings, shown in Fig.~\ref{fig:lgmexp}b, to study the influence of the type-I S$_{1/2}\ket{F=3}\leftrightarrow$ P$_{3/2}\ket{F' = 4}$ transition on the efficiency of $\Lambda$-GMC.
For carrier detunings $\delta_2$ ranging from $+6\,\Gamma$ to $+12.6\,\Gamma$, the Raman detuning shows three regimes visible in Fig.~\ref{fig:lgmexp}b. Blue-detuned of the Raman resonance, a heating feature is observed.  Red-detuned of the resonance, a cooling feature is observed, and far ($|\delta|>0.1\,\Gamma$) from the resonance, neither substantial heating nor cooling occurs. 
As $\delta_2$ increases, the strength of both the cooling and the heating features decreases, eventually completely fading away at $\delta_2=14.3\,\Gamma$.

We show the best-fit temperatures for various carrier detunings $\delta_2$ at a fixed optimal Raman detuning of $\delta=-0.016\,\Gamma$ in Fig.~\ref{fig:lgmexp}c. To quantify the temperature achieved after the $\Lambda$-GMC, we perform release and recapture measurements  and use a Monte Carlo simulation to fit the temperature \cite{PhysRevA.78.033425}. A minimum temperature of \SI{4.0(2)}{\micro\kelvin} is found at a carrier detuning of $\delta_2=6\,\Gamma$, similar to what has been achieved in $^{87}$Rb \cite{D2LambdaCooling}. The temperature after $\Lambda$-GMC is also fitted with a 2D quantized model detailed in App.~\ref{app:release-recapture} that takes into account the zero-point motion. The quantum model yields an average radial motional occupation $\Bar{n}\approx0.7$.

\begin{widetext2}
    \begin{minipage}[b]{\linewidth}
        \begin{figure}[H]
            \centering
            \includegraphics[width=0.67\linewidth]{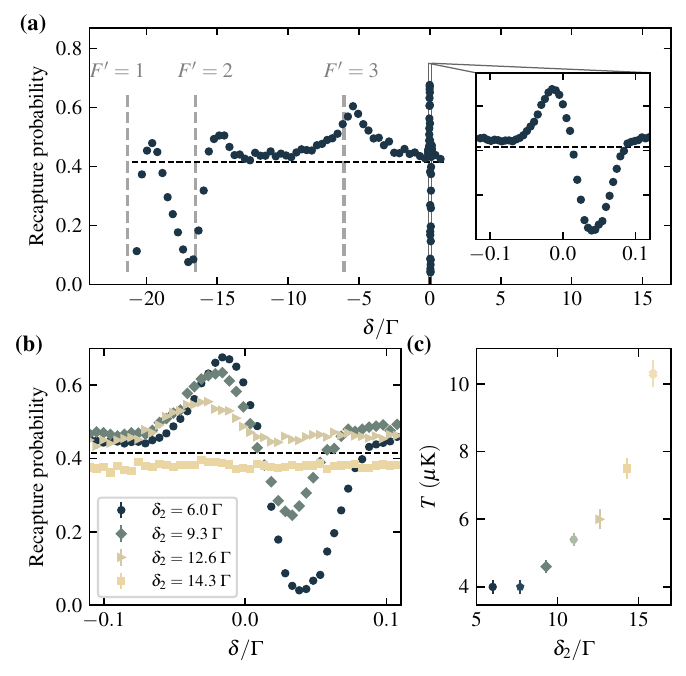}
    \caption{\textbf{$\Lambda$-GMC and heating features in $^{85}$Rb. (a)} Recapture probability after the traps are turned off for \SI{40}{\micro\second} as a function of the Raman detuning $\delta$ for carrier detuning $\delta_2=6\,\Gamma$. Approximate 5$^2$S$_{1/2}$~$\ket{F = 2}$~$\leftrightarrow$~5$^{2}$P$_{3/2}$~$\ket{F' = X}$ resonances are indicated by the vertical lines for $X\in\{1,2,3\}$. The probability of recapture without $\Lambda$-GMC is indicated by the horizontal dashed black line. The inset depicts the narrow cooling/heating feature around the Raman resonance. Error bars (not visible) represent one standard error of the array mean. \textbf{(b)} Release-recapture probability for a fixed $\SI{40}{\micro\second}$ trap off time as a function of Raman detuning $\delta$ for varying $\delta_2$. \textbf{(c)} Temperature as a function of $\delta_2$ after an optimal $\Lambda$-GMC pulse with Raman detuning $\delta=-0.016\,\Gamma$. Temperatures are extracted from Monte Carlo simulations to release-recapture curves, and error bars are derived from the $\chi^2$ value of the Monte Carlo simulations.}
            \label{fig:lgmexp}
        \end{figure}    
    \end{minipage}
\end{widetext2}

\section{Four-level Cooling model}
\subsection{Theory}
\label{sec:Model}
We numerically model the $\Lambda$-enhanced gray molasses cooling force using a four-level system, motivated by the strong dependence on temperature as a function of the detuning $\delta_2$ observed in Sec.~\ref{sec:ExpResults}. The $\Lambda$-GMC force has been studied using the semi-classical force approach in a three-level system \cite{Kosachiov:97}, the wave function Monte-Carlo approach \cite{Molmer:93}, or more recently in atomic tweezers that include trapping potentials \cite{PhysRevA.110.043116}. In our study,  we extend the one-dimensional model \cite{Kosachiov:97, PhysRevA.50.1508} based on the semi-classical force approach to a four-level system featuring a second excited state type-I transition. Although we specifically investigate $^{85}$Rb, this approach is also generally valid for different atomic species and can also be used to model $\Lambda$-systems in lithium atoms \cite{PhysRevA.87.063411}, $^{39}$K \cite{PhysRevA.88.053407}, $^{87}$Rb \cite{D2LambdaCooling}, among others.  

The model includes the ground states $F=2,3$ and the excited states $F'= 3,4$ of $^{85}$Rb, labeled $\ket{1}, \ket{2}, \ket{3}$ and $\ket{4}$ in Fig.~\ref{fig:energydiagramexp}, respectively. A $\Lambda$ configuration is formed between the $\ket{1}\leftrightarrow\ket{3}\leftrightarrow\ket{2}$ states. The optical transitions are driven by two polarization standing waves with Rabi frequencies $\Omega_1$ and $\Omega_2$, at similar frequencies $\omega_1\approx\omega_2=kc$, with wavenumber $k$, and relative phase difference $\phi$. As in previous successful models \cite{Kosachiov:97, PhysRevA.50.1508}, the laser beams are modeled as independent intensity standing waves along each leg of the $\Lambda$-system. The $\ket{1}\leftrightarrow\ket{4}\leftrightarrow\ket{2}$ transitions are not part of the $\Lambda$ configuration, as the $\ket{1}\rightarrow\ket{4}$ transition is dipole-forbidden by selection rules. However, the addressing laser can still induce off-resonant scattering to state $\ket{4}$. Consequently, the $\ket{2}\rightarrow\ket{4}$ transition is treated as a spatially constant coupling. 

The numerical method relies on a Fourier expansion of the density matrix elements as \cite{PhysRevA.50.1508}

\begin{equation}
    \rho_{ij}=\sum_{n=-\infty}^{\infty}\rho_{ij}^{(n)}e^{inkz},
    \label{eq:fourierharmonics}
\end{equation}
with spatial coordinate $z$ and fourier index $n$. Since the solution is obtained through a recurrence equation (App.~\ref{app:model}), a spatially constant coupling without complex exponential in the Hamiltonian cannot be correctly captured. Therefore, we solve the density matrix for two versions of the Hamiltonian, one with a right-traveling wave $\tilde{\Omega}_2(+z) =~\sqrt{2}\Omega_{2}e^{+ikz}$, and one with a left-traveling wave $\tilde{\Omega}_2(-z) = \sqrt{2}\Omega_{2}e^{-ikz}$, which correspond to the $H^{(+)}$ and $H^{(-)}$ Hamiltonian, respectively, defined as

\begin{align}
    H^{(\pm)} &= \hbar \Omega_1 \cos{(kz)}\left(\ket{1}\bra{3} + h.c.\right)\\
    &\quad+ \hbar\Omega_2\cos{(kz+\phi)}\left(\ket{2}\bra{3} + h.c.\right)\\
    &\quad+ \hbar \tfrac{1}{2}(\tilde{\Omega}_2(\pm z)\ket{2}\bra{4}+h.c.)\\
    &\quad+ \hbar \delta_2 \ket{2}\bra{2} + \hbar\delta_1\ket{1}\bra{1} + \hbar\delta_4\ket{4}\bra{4}.
    \label{eq: Hamiltonian}
\end{align}

The force is then averaged over the two individual solutions. A similar averaging procedure is performed in $^{39}$K, where Nath \textit{et al.} add two separate three-level $\Lambda$-systems to their model \cite{PhysRevA.88.053407}.

To compute the cooling dynamics of the atom, the force $F^{(\pm)}(z,v)$ is solved from Eq.~\eqref{eq: Hamiltonian} as the quantum average of the gradient of the potential, $F^{(\pm)}(z,v) = -\mathrm{Tr}[\rho \nabla H^{(\pm)}]$. A more detailed implementation of this continued fractions method for the four-level system is given in App.~\ref{app:model}. Since the typical displacement between the atoms $\Delta z \gg \lambda$, we can limit ourselves to the force averaged over a wavelength

\begin{equation}
    F^{(\pm)}(v) = \frac{1}{\lambda}\int_{0}^{\lambda}F^{(\pm)}(z,v)\mathrm{d}z.
    \label{eq:avgforce}
\end{equation}
The average force from Eq.~\eqref{eq:avgforce} over one wavelength can be rewritten using the Fourier substitution as

\begin{align}
    F^{(\pm)}(v) =& -\frac{i\hbar k}{2}\left[\Omega_1\left(\rho_{31}^{(-1)}+\rho_{13}^{(-1)}-\rho_{31}^{(1)}-\rho_{31}^{(1)}\right)\right.\\    &+\left.\Omega_2\left((\rho_{32}^{(-1)}+\rho_{23}^{(-1)})e^{i\phi}-(\rho_{32}^{(1)}+\rho_{23}^{(1)})e^{-i\phi}\right.\right.\\
    &\left.\left.\pm\sqrt{2}\rho_{42}^{(\mp1)}\mp\sqrt{2}\rho_{24}^{(\pm1)}\right)\right].
    \label{eq:forcesubs}
\end{align}
Although individual forces from Eq.~\eqref{eq:forcesubs} are not balanced and have a zero crossing at finite velocity $v$, their combined force satisfies $F(v=0)=(F^{(+)}(0)+F^{(-)}(0))/2=0$, consistent with a zero crossing at zero velocity. 

The net effect is that the $\Lambda$-system drives the familiar sub-Doppler cooling mechanism based on coherent population trapping at the Raman resonance \cite{PhysRevLett.61.826}, while the additional excited-state coupling to $\ket{4}$ provides position-independent off-resonant scattering. In principle, the resulting force can then be combined with a diffusion coefficient to obtain an estimate of the equilibrium temperature. However, from our simplified approximation to the total atom dynamics, we do not expect a reliable steady-state temperature to be given. 

We restrict the results to the force-velocity relation, from which we derive the friction coefficient $\alpha$, and analyze the photon scattering rates $\Gamma \rho_{33}$ and $\Gamma\rho_{44}$ to reveal the underlying physical mechanisms. The friction coefficient $\alpha$ is obtained in the low-velocity region ($v\ll \Gamma/k$) as a linear fit from

\begin{equation}
    F(v) = -\alpha v,
    \label{eq:def_alpha}
\end{equation}
where $\alpha>0 (<0)$ indicates cooling (heating). If we assume that the phase $\phi$ between the two transitions fluctuates randomly on fast timescales, the phase-averaged force can be calculated from Eq.~\eqref{eq:def_alpha} as

\begin{equation}
    \langle F(v)\rangle_{\phi} = \frac{1}{2\pi}\int_{0}^{2\pi}F(v)\mathrm{d}\phi.
    \label{eq:force-phase-avg}
\end{equation}
\subsection{Numerical Results}
The phase-averaged force defined in Eq.~\eqref{eq:force-phase-avg} is plotted in Fig. \ref{fig:lgmsim}a at the Raman condition where $\delta=0$. For small velocities $|kv/\Gamma |\lesssim 0.01$, cooling is observed, as indicated by the negative slope, and the cooling efficiency is reduced for larger carrier detuning $\delta_2$. The phase-averaged force agrees with existing three-level calculations for $\delta_2\ll\delta_4$ \cite{PhysRevA.87.063411}, and is plotted for different $\delta$ around the Raman resonance condition in App.~\ref{app:model}.

In Fig. \ref{fig:lgmsim}b, the friction coefficient $\alpha$ is plotted as a function of the repumper detuning $\delta_1$ for various fixed carrier detunings $\delta_2$. The Raman resonance can be observed in the curves at $\delta=\delta_1-\delta_2=0$. Upon scanning the repumper detuning, $\delta_1$, first a narrow cooling peak is visible at the Raman resonance, followed by a heating feature. By moving the Raman resonance closer to the $\ket{4}$ level when increasing $\delta_2$, both the cooling and heating efficiency reduce, until they are barely resolved at $\delta_2=17\,\Gamma$. We attribute this reduction to a combination of decreased Raman coupling between the two ground states, scaling as $\Omega_1\Omega_2/\delta_2$, and increased off-resonant scattering to state $\ket{4}$ leading to \lq quasi-dark state' cooling. 

\begin{widetext2}
    \begin{minipage}[b]{\linewidth}
        \begin{figure}[H]
            \centering
            \includegraphics[width=0.8\linewidth]{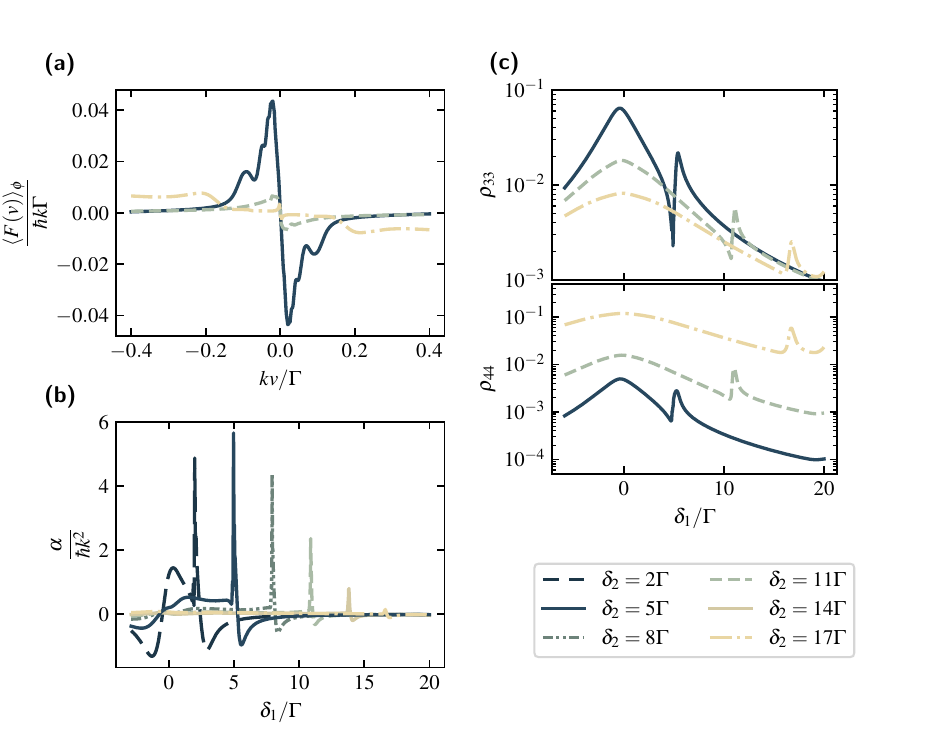}
    \caption{\textbf{Simulation of cooling mechanisms around the Raman resonance. }\textbf{ (a)} Normalized phase-averaged force $\langle F(v)\rangle_{\phi}$ defined in Eq. \eqref{eq:force-phase-avg} at the Raman resonance ($\delta$ = 0) as a function of normalized velocity for $\Omega_1 = 0.58\,\Gamma$, $\Omega_2 = 1.63\,\Gamma$, $\delta_4=19.9\,\Gamma$, and varying carrier detuning $\delta_2$. The cooling is most efficient at $\delta_2=\delta_1=5\,\Gamma$, and the phase-averaged force is presented for different $\delta$ around the Raman resonance in the App~\ref{app:model}. \textbf{(b)} Friction coefficient $\alpha$ as function of repumper detuning $\delta_1$ for varying carrier detuning $\delta_2$ using the parameters $\Omega_1 = 0.58\,\Gamma$, $\Omega_2 = 1.63\,\Gamma$, $\delta_4=19.9\,\Gamma$, and phase $\phi =0$. \textbf{(c)} Population $\rho_{33}$ and $\rho_{44}$ averaged over relative phases $\phi$ as a function of detuning $\delta_1$. Using standard parameters from (a,b) and $kv/\Gamma = 0.01$.}
            \label{fig:lgmsim}
        \end{figure}    
    \end{minipage}
\end{widetext2}

The populations of state $\ket{3}$ and $\ket{4}$ provide more insight into the cooling dynamics at the Raman resonance, as shown in Fig.~\ref{fig:lgmsim}c. Here, the populations $\rho_{33}$ and $\rho_{44}$ are calculated using the continued fractions method, resulting in the scattering rates $\Gamma\rho_{33}$ and $\Gamma\rho_{44}$. For detunings $\delta_2=\{5,11\}\,\Gamma$, the populations exhibit a clear minimum at the Raman resonance, consistent with the suppression of excitations from the dark state \cite{PhysRevResearch.6.023154}. At $\delta_2=11\,\Gamma$, roughly $\delta_4/2$, a cross-over region is visible where the populations $\rho_{33}$ and $\rho_{44}$ are comparable in magnitude.  
At larger detunings, $\delta_2=17\,\Gamma$, the characteristic dispersive feature no longer exhibits a pronounced minimum but instead reveals only a small heating peak, and $\rho_{44}$ dominates over $\rho_{33}$. In this regime, the reduction of scattering at the Raman resonance is absent, thereby eliminating the velocity selective cooling mechanism.  

\section{Comparison Simulation and Experiment}
\label{sec:discussion}

We experimentally studied the reduction and eventual disappearance of the characteristic $\Lambda$-GMC feature around the Raman resonance as the carrier detuning $\delta_2$ increases. Similar to observations in $^{87}$Rb \cite{D2LambdaCooling}, as the detuning of the type-I transition is reduced, the heating feature reduces in amplitude. In $^{85}$Rb we observe that the maximum cooling and heating occurs at $\delta_2=6\,\Gamma$, which is the minimum detuning that we are able to investigate in our experiment, and both the heating and cooling are reduced with increasing $\delta_2$. At $\delta_2=14.3\, \Gamma$, which is $5.5\,\Gamma$ red-detuned of the type-I transition, no heating or cooling is observed near the Raman resonance. 

The simulation results mirror our experimental findings, where the disappearance of cooling and heating is observed around $\delta_2\approx15\,\Gamma$, as indicated by the force-velocity dependence and the friction coefficient $\alpha$. With increasing $\delta_2$, the cooling force is significantly reduced around the Raman resonance, due to the increase in off-resonant excitation to the $\ket{4}$ state, redistributing the overall scattering rates. This redistribution can be observed in the populations $\rho_{33}$ and $\rho_{44}$ and spans several orders of magnitude, giving insight into the competing roles of the two excited states in the heating and cooling mechanisms. 

The loss of cooling efficiency in simulations at carrier detuning $\delta_2=17\,\Gamma$ and Raman detuning $\delta=0\,\Gamma$ is specific to four-level systems with intermediate excited-state hyperfine splitting and therefore applies to $^{85}$Rb. This contrasts with $^{87}$Rb, which has a large hyperfine splitting, or three-level systems, where reduced cooling persists at the same absolute detuning. In general, a reduced two-photon coupling is always observed for three- or four-level systems with increasing the carrier detuning $\delta_2$ at the Raman resonance condition.

\section{Extended $T_2^*$ of Hyperfine Qubits}
To investigate the coherence properties of the hyperfine clock qubit, defined as $\ket{0}:=\ket{F=2,m_F=0}$ and $\ket{1}:=\ket{F=3, m_F=0}$, we measure the $T_2^*$ coherence time including the $\Lambda$-GMC pulse. The measurement is executed with a Ramsey-type pulse sequence using $\SI{2}{\kilo\hertz}$ detuned microwaves on the $\ket{0}\leftrightarrow\ket{1}$ transition at $\SI{3}{\giga\hertz}$. 

\begin{figure}
    \centering
    \includegraphics[width=0.9\linewidth]{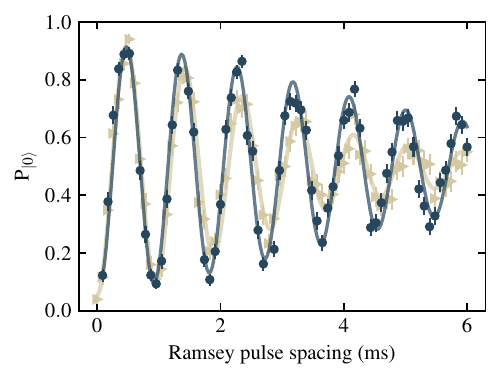}
    \caption{\textbf{Damped Ramsey oscillation including fit for a single array site.} The circular data points (dark blue) represent the Ramsey signal after $\Lambda$-GMC, and triangular data points (green) represent the Ramsey signal without $\Lambda$-GMC. Error bars are statistical and represent one standard error of the mean. }
    \label{fig:RamseyT2LGM}
\end{figure}

In Fig.~\ref{fig:RamseyT2LGM} the Ramsey oscillation is shown with and without a $\Lambda$-GMC pulse ($\delta_2=6$~$\Gamma$, $\delta = -0.016$~$\Gamma$), and corresponds to the ensemble average coherence time $\left<T_2^*\right>$ of \SI{5.3(2)}{\milli\second} and \SI{3.4(1)}{\milli\second} respectively. We analyze the array site individually to prevent trap depth inhomogeneities from dominating the decay of the array averaged signal. The fit to the measurements is of the form 

\begin{equation}
P_{\ket{0}} = A+\alpha\left(t,T_2^*\right)B\cos{\left(t\Delta+ \kappa\left(t,T_2^*\right)+\phi\right)}, \label{eqn:T2Fit}
\end{equation}
where $\Delta$ is the microwave detuning and $\alpha\left(t,T_2^*\right)$ and $\kappa\left(t,T_2^*\right)$ are the time-dependent oscillation amplitude and phase shift, respectively, and $A$ and $B$ are fit parameters \cite{PhysRevA.72.023406}. Due to heating during the state preparation procedure (see App.~\ref{app:release-recapture}), we expect that a further performance increase in coherence times can be expected using alternative state preparation methods \cite{66s8-jj18, lukin_2025_state_prep}.

There exist also other dephasing mechanisms in addition to the finite atom temperature, and the combined effect leads to the total measured $T_2^*$ times. One of such dephasing mechanisms can be related to magnetic field noise, which we have previously estimated to be $\delta B< 1.8$\,mG \cite{vanrensburg2025fidelityrelationsarrayneutral}. This level of magnetic field noise would limit the $T_2^*$ time of the clock ($m_F=0$) qubit to \SI{55}{\milli\second} when considering the quadratic Zeeman effect. A second effect could be fluctuations in the depths of the tweezer traps. We stabilize the tweezer intensity with a PI controller \cite{preuschoff_digital_2020} using a photodiode located before the SLM. Moreover, the beam after the SLM may suffer from SLM phase flicker at \SI{1.4}{\kilo\hertz}.

\section{Discussion and Outlook}
We have experimentally demonstrated $\Lambda$-GMC of single $^{85}$Rb atoms in optical tweezers on the D$_2$ line, reaching temperatures of $\SI{4.0(2)}{\micro\kelvin}$ with the recommended operating parameters of a carrier detuning $\delta_2=6\,\Gamma$ and Raman detuning of $\delta=-0.016\,\Gamma$. Based on the quantized model (App.~\ref{app:release-recapture}), this corresponds to an average radial motional level of $\Bar{n}\approx 0.7$.  The use of the D$_2$ line is convenient, as a single laser with an EOM sideband can be used to operate a MOT, single-atom imaging, hyperfine qubit state preparation, qubit state detection, and $\Lambda$-GMC. Using MOT beams for $\Lambda$-GMC also eliminates the need to perform any further alignment. The inclusion of $\Lambda$-GMC was shown to extend the $T_2^*$ coherence time of hyperfine qubits in non-magic traps, up to a factor of $1.5$. This demonstration of $\Lambda$-GMC in $^{85}$Rb shows that the technique can be applied to prepare dual-species arrays of $^{85}$Rb and $^{87}$Rb atoms to similarly low temperatures using the $D_2$ line.

We have extended the continued-fractions model to a four-level system in order to evaluate the role of additional excited states, and the approach is readily generalizable to other four-level systems. The calculated force and friction coefficient indicate maximum $\Lambda$-GMC at carrier detuning of $\delta_2=5\,\Gamma$. Furthermore, for carrier detunings approaching the excited state $\ket{4}$, the efficiency of the $\Lambda$-GMC decreases sharply, in agreement with the experimental observations for $^{85}$Rb. For future work, a full three-dimensional calculation in tweezers could be performed that includes all relevant ground and excited states \cite{PhysRevA.110.043116}, providing more precision than the simplified four-level model.

\section*{ACKNOWLEDGEMENTS}
We thank the SrMic team at the University of Amsterdam, the Sr quantum computing team at Eindhoven University of Technology and R.M.P. Teunissen for fruitfull discussions. This research is financially supported by the Dutch Ministry of Economic Affairs and Climate Policy (EZK), as part of the Quantum Delta NL program,  the Horizon Europe programme HORIZON-CL4-2021-DIGITAL-EMERGING-01-30 via the project 101070144 (EuRyQa), and by the Netherlands Organisation for Scientific Research (NWO) under Grant No.\ 680.92.18.05. 

\section*{AUTHOR CONTRIBUTIONS}

D.J.v.R conceptualized the experiment, collected and analyzed the experimental data, and together with R.V. drafted the manuscript. R.V. developed the model, performed the numerical calculations, and interpreted their results. D.J.v.R., R.V., Y.v.d.W. and J.d.P.M participated in the design, construction and characterization of the experimental setup, M.V. performed the analysis for the quantized release-recapture model. E.V., R.L. and S.K. supervised the work. 

\section*{COMPETING INTERESTS}
The authors declare no competing interests.

\section*{DATA AND CODE  AVAILABILITY}
The data and code that support the findings of this study are available in the 4TU database \cite{4TU}. Additional data is available from the corresponding author upon reasonable request.

\newpage
\bibliography{Bibliography.bib}

@article{D2LambdaCooling,
    author = {Sara Rosi and Alessia Burchianti and Stefano Conclave and Devang S. Naik and Giacomo Roati and Chiara Fort and Francesco Minardi},
    title = {$\Lambda$-enhanced grey molasses on the D2 transition of Rubidium-87 atoms},
    journal = {Scientific Reports},
    volume = {8},
    pages = {1301},
    year = {2013},
    doi = {https://www.nature.com/articles/s41598-018-19814-z#citeas}
}

@article{PhysRevA.50.1508,
  title = {Sub-Doppler cooling of three-level \ensuremath{\Lambda} atoms in space-shifted standing light waves},
  author = {Kosachiov, D. and Rozhdestvensky, Yu. and Olsen, M. and Plimak, L. and Walls, D. F.},
  journal = {Phys. Rev. A},
  volume = {50},
  issue = {2},
  pages = {1508--1512},
  numpages = {0},
  year = {1994},
  month = {Aug},
  publisher = {American Physical Society},
  doi = {10.1103/PhysRevA.50.1508},
  url = {https://link.aps.org/doi/10.1103/PhysRevA.50.1508}
}

@article{PhysRevA.88.053407,
  title = {Quantum-interference-enhanced deep sub-Doppler cooling of ${}^{39}$K atoms in gray molasses},
  author = {Nath, Dipankar and Easwaran, R Kollengode and Rajalakshmi, G. and Unnikrishnan, C. S.},
  journal = {Phys. Rev. A},
  volume = {88},
  issue = {5},
  pages = {053407},
  numpages = {7},
  year = {2013},
  month = {Nov},
  publisher = {American Physical Society},
  doi = {10.1103/PhysRevA.88.053407},
  url = {https://link.aps.org/doi/10.1103/PhysRevA.88.053407}
}

@article{PhysRevLett.61.826,
  title = {Laser Cooling below the One-Photon Recoil Energy by Velocity-Selective Coherent Population Trapping},
  author = {Aspect, A. and Arimondo, E. and Kaiser, R. and Vansteenkiste, N. and Cohen-Tannoudji, C.},
  journal = {Phys. Rev. Lett.},
  volume = {61},
  issue = {7},
  pages = {826--829},
  numpages = {0},
  year = {1988},
  month = {Aug},
  publisher = {American Physical Society},
  doi = {10.1103/PhysRevLett.61.826},
  url = {https://link.aps.org/doi/10.1103/PhysRevLett.61.826}
}

@article{Kosachiov:97,
author = {D. V. Kosachiov and Yu. V. Rozhdestvensky and G. Nienhuis},
journal = {J. Opt. Soc. Am. B},
number = {3},
pages = {535--543},
publisher = {Optica Publishing Group},
title = {Laser cooling of three-level atoms in two standing waves},
volume = {14},
month = {Mar},
year = {1997},
url = {https://opg.optica.org/josab/abstract.cfm?URI=josab-14-3-535},
doi = {10.1364/JOSAB.14.000535},
}

@article{Molmer:93,
author = {Klaus M{\o}lmer and Yvan Castin and Jean Dalibard},
journal = {J. Opt. Soc. Am. B},
number = {3},
pages = {524--538},
publisher = {Optica Publishing Group},
title = {Monte Carlo wave-function method in quantum optics},
volume = {10},
month = {Mar},
year = {1993},
url = {https://opg.optica.org/josab/abstract.cfm?URI=josab-10-3-524},
doi = {10.1364/JOSAB.10.000524},
}

@article{PhysRevA.87.063411,
  title = {$\ensuremath{\Lambda}$-enhanced sub-Doppler cooling of lithium atoms in ${D}_{1}$ gray molasses},
  author = {Grier, Andrew T. and Ferrier-Barbut, Igor and Rem, Benno S. and Delehaye, Marion and Khaykovich, Lev and Chevy, Fr\'ed\'eric and Salomon, Christophe},
  journal = {Phys. Rev. A},
  volume = {87},
  issue = {6},
  pages = {063411},
  numpages = {8},
  year = {2013},
  month = {Jun},
  publisher = {American Physical Society},
  doi = {10.1103/PhysRevA.87.063411},
  url = {https://link.aps.org/doi/10.1103/PhysRevA.87.063411}
}

@article{PhysRevResearch.6.023154,
  title = {Fano resonance in excitation spectroscopy and cooling of an optically trapped single atom},
  author = {Chow, Chang Hoong and Ng, Boon Long and Prakash, Vindhiya and Kurtsiefer, Christian},
  journal = {Phys. Rev. Res.},
  volume = {6},
  issue = {2},
  pages = {023154},
  numpages = {8},
  year = {2024},
  month = {May},
  publisher = {American Physical Society},
  doi = {10.1103/PhysRevResearch.6.023154},
  url = {https://link.aps.org/doi/10.1103/PhysRevResearch.6.023154}
}

@article{66s8-jj18,
  title = {Universal Neutral-Atom Quantum Computer with Individual Optical Addressing and Nondestructive Readout},
  author = {Radnaev, A.G. and Chung, W.C. and Cole, D.C. and Mason, D. and Ballance, T.G. and others},
  journal = {PRX Quantum},
  volume = {6},
  issue = {3},
  pages = {030334},
  numpages = {20},
  year = {2025},
  month = {Aug},
  publisher = {American Physical Society},
  doi = {10.1103/66s8-jj18},
  url = {https://link.aps.org/doi/10.1103/66s8-jj18}
}

@article{10.1116/5.0036562,
    author = {Morgado, M. and Whitlock, S.},
    title = {Quantum simulation and computing with Rydberg-interacting qubits},
    journal = {AVS Quantum Science},
    volume = {3},
    number = {2},
    pages = {023501},
    year = {2021},
    month = {05},
    issn = {2639-0213},
    doi = {10.1116/5.0036562},
    url = {https://doi.org/10.1116/5.0036562},
}

@article{PhysRevA.110.043116,
  title = {Generalized theory for optical cooling of a trapped atom with spin},
  author = {Phatak, Saumitra S. and Blodgett, Karl N. and Peana, David and Chen, Meng Raymond and Hood, Jonathan D.},
  journal = {Phys. Rev. A},
  volume = {110},
  issue = {4},
  pages = {043116},
  numpages = {20},
  year = {2024},
  month = {Oct},
  publisher = {American Physical Society},
  doi = {10.1103/PhysRevA.110.043116},
  url = {https://link.aps.org/doi/10.1103/PhysRevA.110.043116}
}

@misc{vanrensburg2025fidelityrelationsarrayneutral,
      title={Fidelity Relations in an Array of Neutral Atom Qubits -- Experimental Validation of Control Noise}, 
      author={Deon Janse van Rensburg and Robert de Keijzer and Rogier Venderbosch and Yuri van der Werf and Jesus del Pozo Mellado and Rianne Lous and Edgar Vredenbregt and Servaas Kokkelmans},
      year={2025},
      eprint={2506.16974},
      archivePrefix={arXiv},
      primaryClass={quant-ph},
      url={https://arxiv.org/abs/2506.16974}, 
}

@article{magictrapping,
  title = {Gate fidelity, dephasing, and ‘magic’ trapping of optically trapped neutral atom},
  author = {Yang, Pengfei and Li, Gang and Wang, Zhihui and Zhang, Pengfei and Zhang, Tiancai},
  journal = {New J. Phys},
  volume = {24},
  pages = {083028},
  year = {2022},
  month = {August},
  publisher = {Physics and Deutsche Physikalische Gesellschaft},
  doi = {10.1088/1367-2630/ac87ca},
  url = {https://iopscience.iop.org/article/10.1088/1367-2630/ac87ca#njpac87cas5}
}

@incollection{GRIMM200095,
title = {Optical Dipole Traps for Neutral Atoms},
editor = {Benjamin Bederson and Herbert Walther},
series = {Advances In Atomic, Molecular, and Optical Physics},
publisher = {Academic Press},
volume = {42},
pages = {95-170},
year = {2000},
issn = {1049-250X},
doi = {https://doi.org/10.1016/S1049-250X(08)60186-X},
url = {https://www.sciencedirect.com/science/article/pii/S1049250X0860186X},
author = {Rudolf Grimm and Matthias Weidemüller and Yurii B. Ovchinnikov}
}

@article{Salomon_2013,
doi = {10.1209/0295-5075/104/63002},
url = {https://dx.doi.org/10.1209/0295-5075/104/63002},
year = {2014},
month = {jan},
publisher = {EDP Sciences, IOP Publishing and Società Italiana di Fisica},
volume = {104},
number = {6},
pages = {63002},
author = {Salomon, G. and Fouché, L. and Wang, P. and Aspect, A. and Bouyer, P. and Bourdel, T.},
title = {Gray-molasses cooling of 39K to a high phase-space density},
journal = {Europhysics Letters}
}

@article{rio_fernandes_sub-doppler_2012,
	title = {Sub-{Doppler} laser cooling of fermionic {40K} atoms in three-dimensional gray optical molasses},
	volume = {100},
	url = {https://dx.doi.org/10.1209/0295-5075/100/63001},
	doi = {10.1209/0295-5075/100/63001},
	number = {6},
	journal = {Europhysics Letters},
	author = {Rio Fernandes, D. and Sievers, F. and Kretzschmar, N. and Wu, S. and Salomon, C. and Chevy, F.},
	month = dec,
	year = {2012},
	pages = {63001},
}

@article{PhysRevA.94.033408,
  title = {Production of large $^{41}\mathrm{K}$ Bose-Einstein condensates using ${D}_{1}$ gray molasses},
  author = {Chen, Hao-Ze and Yao, Xing-Can and Wu, Yu-Ping and Liu, Xiang-Pei and Wang, Xiao-Qiong and Wang, Yu-Xuan and Chen, Yu-Ao and Pan, Jian-Wei},
  journal = {Phys. Rev. A},
  volume = {94},
  issue = {3},
  pages = {033408},
  numpages = {6},
  year = {2016},
  month = {Sep},
  publisher = {American Physical Society},
  doi = {10.1103/PhysRevA.94.033408},
  url = {https://link.aps.org/doi/10.1103/PhysRevA.94.033408}
}

@article{PhysRevA.93.023421,
  title = {Sub-Doppler cooling of sodium atoms in gray molasses},
  author = {Colzi, Giacomo and Durastante, Gianmaria and Fava, Eleonora and Serafini, Simone and Lamporesi, Giacomo and Ferrari, Gabriele},
  journal = {Phys. Rev. A},
  volume = {93},
  issue = {2},
  pages = {023421},
  numpages = {6},
  year = {2016},
  month = {Feb},
  publisher = {American Physical Society},
  doi = {10.1103/PhysRevA.93.023421},
  url = {https://link.aps.org/doi/10.1103/PhysRevA.93.023421}
}

@article{PhysRevA.98.033419,
  title = {$\mathrm{\ensuremath{\Lambda}}$-enhanced gray-molasses cooling of cesium atoms on the ${D}_{2}$ line},
  author = {Hsiao, Ya-Fen and Lin, Yu-Ju and Chen, Ying-Cheng},
  journal = {Phys. Rev. A},
  volume = {98},
  issue = {3},
  pages = {033419},
  numpages = {6},
  year = {2018},
  month = {Sep},
  publisher = {American Physical Society},
  doi = {10.1103/PhysRevA.98.033419},
  url = {https://link.aps.org/doi/10.1103/PhysRevA.98.033419}
}

@article{bruce_sub-doppler_2017,
	title = {Sub-{Doppler} laser cooling of {40K} with {Raman} gray molasses on the D$_2$ line},
	volume = {50},
	url = {https://dx.doi.org/10.1088/1361-6455/aa65ea},
	doi = {10.1088/1361-6455/aa65ea},
	number = {9},
    pages = {095002},
	journal = {Journal of Physics B: Atomic, Molecular and Optical Physics},
	author = {Bruce, G D and Haller, E and Peaudecerf, B and Cotta, D A and Andia, M and others},
	month = apr,
	year = {2017},
}

@article{Shi_2018,
doi = {10.1088/0256-307X/35/12/123701},
url = {https://dx.doi.org/10.1088/0256-307X/35/12/123701},
year = {2018},
month = {dec},
publisher = {Chinese Physical Society and IOP Publishing Ltd},
volume = {35},
number = {12},
pages = {123701},
author = {Shi, Zhenlian and Li, Ziliang and Wang, Pengjun and Meng, Zengming and Huang, Lianghui and Zhang, Jing},
title = {Sub-Doppler Laser Cooling of 23Na in Gray Molasses on the D2 Line},
journal = {Chinese Physics Letters}
}

@article{graham_multi-qubit_2022,
	title = {Multi-qubit entanglement and algorithms on a neutral-atom quantum computer},
	volume = {604},
	optissn = {1476-4687},
	url = {https://doi.org/10.1038/s41586-022-04603-6},
	doi = {10.1038/s41586-022-04603-6},
	number = {7906},
	journal = {Nature},
	author = {Graham, T. M. and Song, Y. and Scott, J. and Poole, C. and Phuttitarn, L. and Jooya, K. and Eichler, P. and others},
	month = apr,
	year = {2022},
	pages = {457--462},
}

@article{PhysRevA.78.033425,
  title = {Energy distribution and cooling of a single atom in an optical tweezer},
  author = {Tuchendler, C. and Lance, A. M. and Browaeys, A. and Sortais, Y. R. P. and Grangier, P.},
  journal = {Phys. Rev. A},
  volume = {78},
  issue = {3},
  pages = {033425},
  numpages = {9},
  year = {2008},
  month = {Sep},
  publisher = {American Physical Society},
  doi = {10.1103/PhysRevA.78.033425},
  url = {https://link.aps.org/doi/10.1103/PhysRevA.78.033425}
}

@article{manetsch_tweezer_2025,
	title = {A tweezer array with 6100 highly coherent atomic qubits},
	issn = {1476-4687},
	url = {https://doi.org/10.1038/s41586-025-09641-4},
	doi = {10.1038/s41586-025-09641-4},
	journal = {Nature},
    volume = {647},
    pages = {60--67},
	author = {Manetsch, Hannah J. and Nomura, Gyohei and Bataille, Elie and Lv, Xudong and Leung, Kon H. and Endres, Manuel},
	month = sep,
	year = {2025},
}

@article{lukin_2025_state_prep,
	title = {Continuous operation of a coherent 3,000-qubit system},
	issn = {1476-4687},
	url = {https://doi.org/10.1038/s41586-025-09596-6},
	doi = {10.1038/s41586-025-09596-6},
	journal = {Nature},
    volume = {646},
    pages = {1075--1080},
	author = {Chiu, N.C. and Trapp, E.C. and Guo, J. and others},
	month = sep,
	year = {2025},
}

@article{PhysRevLett.120.083201,
  title = {Blue-Detuned Magneto-Optical Trap},
  author = {Jarvis, K. N. and Devlin, J. A. and Wall, T. E. and Sauer, B. E. and Tarbutt, M. R.},
  journal = {Phys. Rev. Lett.},
  volume = {120},
  issue = {8},
  pages = {083201},
  numpages = {5},
  year = {2018},
  month = {Feb},
  publisher = {American Physical Society},
  doi = {10.1103/PhysRevLett.120.083201},
  url = {https://link.aps.org/doi/10.1103/PhysRevLett.120.083201}
}

@article{preuschoff_digital_2020,
	title = {Digital laser frequency and intensity stabilization based on the {STEMlab} platform (originally {Red} {Pitaya})},
	volume = {91},
	issn = {0034-6748},
	url = {https://doi.org/10.1063/5.0009524},
	doi = {10.1063/5.0009524},
	number = {8},
	journal = {Review of Scientific Instruments},
	author = {Preuschoff, T. and Schlosser, M. and Birkl, G.},
	month = aug,
	year = {2020},
	pages = {083001},
}

@article{PhysRevA.72.023406,
  title = {Analysis of dephasing mechanisms in a standing-wave dipole trap},
  author = {Kuhr, S. and Alt, W. and Schrader, D. and Dotsenko, I. and Miroshnychenko, Y. and Rauschenbeutel, A. and Meschede, D.},
  journal = {Phys. Rev. A},
  volume = {72},
  issue = {2},
  pages = {023406},
  numpages = {12},
  year = {2005},
  month = {Aug},
  publisher = {American Physical Society},
  doi = {10.1103/PhysRevA.72.023406},
  url = {https://link.aps.org/doi/10.1103/PhysRevA.72.023406}
}

@phdthesis{deonThesis,
title = "Developing a Quantum Computing Platform Based on Single 85Rb Atoms",
author = "Janse van Rensburg, Deon Anton",
note = "Proefschrift.",
year = "2025",
month = oct,
day = "28",
language = "English",
isbn = "978-90-386-6496-5",
publisher = "Eindhoven University of Technology",
type = "Phd Thesis",
school = "Applied Physics and Science Education",
}

@article{Hlzl2023,
   author = {C. Hölzl and A. Götzelmann and M. Wirth and M. S. Safronova and S. Weber and F. Meinert},
   doi = {10.1103/PhysRevResearch.5.033093},
   issn = {26431564},
   issue = {3},
   journal = {Physical Review Research},
   month = {7},
   publisher = {American Physical Society},
   title = {Motional ground-state cooling of single atoms in state-dependent optical tweezers},
   volume = {5},
   year = {2023}
}

@misc{4TU,
  doi = {10.4121/8b5272c5-0457-4175-80e4-7c51a9de70c2},
  author = {Deon Janse van Rensburg and Rogier Venderbosch and Yuri van der Werf and Jesus del Pozo Mellado and Marijn Venderbosch and Rianne Lous and Edgar Vredenbregt and Servaas Kokkelmans},
  title = {Data and Code for \lq\ensuremath{\Lambda}-enhanced Gray Molasses Cooling of \ensuremath{^{85}}Rb Atoms in Tweezers Using the ${D}_{2}$ Line"},
  publisher = {4TU.ResearchData},
  year = {2025},
  copyright = {MIT},
}

@article{Manzano2020,
   author = {Daniel Manzano},
   doi = {10.1063/1.5115323/1021638},
   issn = {21583226},
   issue = {2},
   journal = {AIP Advances},
   month = {2},
   publisher = {American Institute of Physics Inc.},
   title = {A short introduction to the Lindblad master equation},
   volume = {10},
   url = {/aip/adv/article/10/2/025106/1021638/A-short-introduction-to-the-Lindblad-master},
   year = {2020}
}

@article{DeKeijzer2023,
   author = {R. J.P.T. De Keijzer and O. Tse and S. J.J.M.F. Kokkelmans},
   doi = {10.1103/PhysRevA.108.023122},
   issn = {24699934},
   issue = {2},
   journal = {Physical Review A},
   month = {8},
   publisher = {American Physical Society},
   title = {Recapture probability for antitrapped Rydberg states in optical tweezers},
   volume = {108},
   url = {http://arxiv.org/abs/2303.08783 http://dx.doi.org/10.1103/PhysRevA.108.023122},
   year = {2023}
}

@article{evered_high-fidelity_2023,
	title = {High-fidelity parallel entangling gates on a neutral-atom quantum computer},
	volume = {622},
	issn = {1476-4687},
	url = {https://doi.org/10.1038/s41586-023-06481-y},
	doi = {10.1038/s41586-023-06481-y},
	number = {7982},
	journal = {Nature},
	author = {Evered, Simon J. and Bluvstein, Dolev and Kalinowski, Marcin and Ebadi, Sepehr and Manovitz, Tom and others},
	month = oct,
	year = {2023},
	pages = {268--272},
}

@article{RevModPhys.82.2313,
  title = {Quantum information with Rydberg atoms},
  author = {Saffman, M. and Walker, T. G. and M\o{}lmer, K.},
  journal = {Rev. Mod. Phys.},
  volume = {82},
  issue = {3},
  pages = {2313--2363},
  numpages = {0},
  year = {2010},
  month = {Aug},
  publisher = {American Physical Society},
  doi = {10.1103/RevModPhys.82.2313},
  url = {https://link.aps.org/doi/10.1103/RevModPhys.82.2313}
}

@misc{Steck,
      title={Rubidium 85 D Line Data}, 
      author={Daniel A. Steck},
      year={2024},
      month = {May},
      revision = {revision 2.3.3},
      opteprint={2403.12021},
      url={http://steck.us/alkalidata}, 
}

@article{bluvstein_logical_2024,
	title = {Logical quantum processor based on reconfigurable atom arrays},
	volume = {626},
	issn = {1476-4687},
	url = {https://doi.org/10.1038/s41586-023-06927-3},
	doi = {10.1038/s41586-023-06927-3},
	number = {7997},
	journal = {Nature},
	author = {Bluvstein, Dolev and Evered, Simon J. and Geim, Alexandra A. and others},
	month = feb,
	year = {2024},
	pages = {58-65},
}

@article{covey2019,
  title = {2000-{{Times Repeated Imaging}} of {{Strontium Atoms}} in {{Clock-Magic Tweezer Arrays}}},
  author = {Covey, Jacob P. and Madjarov, Ivaylo S. and Cooper, Alexandre and Endres, Manuel},
  year = 2019,
  month = may,
  journal = {Phys. Rev. Lett.},
  volume = {122},
  number = {17},
  pages = {173201},
  issn = {0031-9007, 1079-7114},
  doi = {10.1103/PhysRevLett.122.173201},
  urldate = {2022-09-26},
  langid = {english}
}

\newpage

\appendix

\renewcommand{\thefigure}{A\arabic{figure}}
\setcounter{figure}{0}
  
\section{Release-Recapture Temperature Measurement}
\label{app:release-recapture}
In this section, we provide example release-recapture curves and the method for the temperature data shown in Fig.~\ref{fig:lgmexp}c, which are based on a classical Monte-Carlo analysis \cite{PhysRevA.78.033425}.
The classical release-recapture technique can overestimate the temperature when the probability of occupying the motional ground state is significant \cite{covey2019}.
To more accurately determine the temperature in this regime, we also fit the lowest temperature dataset using a quantized model that accounts for zero-point motion.
This model initializes the atom as a freely expanding thermal state of harmonic oscillator states \cite{Hlzl2023}.
The recapture probability is calculated by computing the overlap between this expanded state and the bound states of a Gaussian potential \cite{DeKeijzer2023}.

\begin{figure}[h!]
    \centering
    \includegraphics[width=\linewidth]{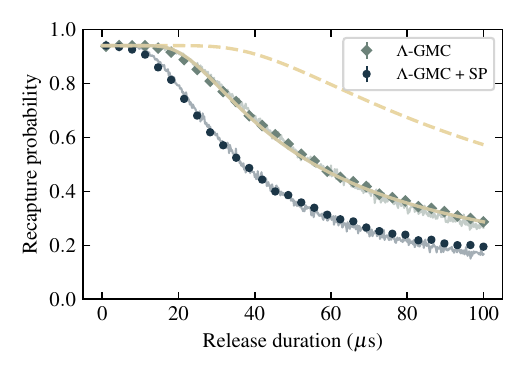}
    \caption{\textbf{Release and recapture measurements.} Curves after $\Lambda$-GMC with and without state preparation (SP) after cooling, including the best fit Monte Carlo simulation curves as faint lines. The yellow-green line shows the fit of the quantum model to the coldest dataset after optimal $\Lambda$-GMC, and the yellow dashed line depicts the radial expansion due to the zero-point motion energy in the trap. Error bars are statistical and represent one standard error of the array mean.
    }
    \label{fig:release-recapture}
\end{figure}

The coldest dataset after optimal $\Lambda$-GMC is fitted with the quantized model in Fig. \ref{fig:release-recapture}, which gives a temperature of $T_{\mathrm{q}}=\SI{2.48(1)}{\micro\kelvin}$, and corresponds to an average radial motional occupation $\Bar{n}\approx 0.7$ and ground state population $p_0\approx0.6$.
The best fit classical temperatures extracted from the Monte Carlo simulations are \SI{4.0(2)}{\micro\kelvin} after $\Lambda$-GMC. Although the classical temperature is higher than the quantized model temperature, we quote the classical temperatures to allow for a more direct comparison with the existing literature. 

After state preparation, which we currently implement based on a microwave-assisted optical pumping type process \cite{deonThesis}, we extract a classical temperature of $\SI{7.6(3)}{\micro\kelvin}$. Our state preparation method thus leads to significant heating of the $\Lambda$-GMC pre-cooled atoms. We anticipate that with millisecond switching of the magnetic field coils alternative state preparation methods can be executed with reduced heating \cite{66s8-jj18}.

\newpage
\section{Simulation model}
\renewcommand{\thefigure}{B\arabic{figure}}
\setcounter{figure}{0}
\label{app:model}

The model for $\Lambda$-enhanced cooling assumes a simplified one-dimensional configuration where the polarization-dependent coupling is introduced as a spatially varying Rabi frequency as in Refs.~\cite{Kosachiov:97, PhysRevA.50.1508}. The extended Hamiltonians for a four-level system defined in Eq. \eqref{eq: Hamiltonian} are then solved using the continued fraction method. To simplify the solution in the steady-state limit, we introduce the hydrodynamic derivative, which is valid under the condition $v\ll \Gamma/k$, as 

\begin{equation}
    \frac{\mathrm{d}}{\mathrm{d}t}=\frac{\partial}{\partial t} +v\frac{\partial}{\partial z}.
    \label{eq:hydrodynamic_derivative}
\end{equation}

Upon substitution in the Lindblad equation \cite{Manzano2020}, the following sets of equations are obtained in the steady-state limit for the Hamiltonians $H^{(\pm)}$ defined in Eq.~\eqref{eq: Hamiltonian} making use of the notation $\tilde{\Omega}_2( z)=\tfrac{\Omega_2}{\sqrt{2}}e^{ikz}$:

\begin{align}
    v \frac{\partial \rho_{11}}{\partial z} &= i\Omega_1(z)(\rho_{13}-\rho_{31}) + \tfrac{4}{9}\Gamma\rho_{33},\\
    v \frac{\partial \rho_{22}}{\partial z} &= i\Omega_2(z)(\rho_{23}-\rho_{32})+i(\tilde{\Omega}_2(\mp z)\rho_{24}-\tilde{\Omega}_2(\pm z)\rho_{42})\\& \quad+ \Gamma(\tfrac{5}{9}\rho_{33} + \rho_{44}),\\ 
    v \frac{\partial \rho_{44}}{\partial z} &= -\Gamma\rho_{44}-i\tilde{\Omega}_2(\mp z)\rho_{24}+i\tilde{\Omega}_2(\pm z)\rho_{42},\\
     v \frac{\partial \rho_{13}}{\partial z} &= (-i\delta_1-\tfrac{1}{2}\Gamma)\rho_{13}  + i\Omega_1(z)(\rho_{11}-\rho_{33})+i\Omega_2(z)\rho_{12},\\
     v \frac{\partial \rho_{14}}{\partial z} &= (i\delta_4-i\delta_1-\tfrac{1}{2}\Gamma)\rho_{14}  -i\Omega_1(z)\rho_{34}+i\tilde{\Omega}_2(\pm z)\rho_{12},\\
     v \frac{\partial \rho_{23}}{\partial z} &= (-i\delta_2-\tfrac{1}{2}\Gamma)\rho_{23}  -i\Omega_2(z)( \rho_{33}-\rho_{22})\\
      &\quad+i\Omega_1(z)\rho_{21}-i\tilde{\Omega}_2(\pm z)\rho_{43},\\
      v \frac{\partial \rho_{24}}{\partial z} &= (i\delta_4-i\delta_2-\tfrac{1}{2}\Gamma)\rho_{24}  -i\Omega_2(z)\rho_{34}\\
    &\quad + i\tilde{\Omega}_2(\pm z)(\rho_{22}-\rho_{44}),\\
     v \frac{\partial \rho_{21}}{\partial z} &= (i\delta_1-i\delta_2)\rho_{21}  -i \Omega_2(z)\rho_{31}\\
    &\quad -i\tilde{\Omega}_2(\pm z)\rho_{41} +i \Omega_1(z)\rho_{23},\\
    v \frac{\partial \rho_{43}}{\partial z} &= (-i\delta_4-\Gamma)\rho_{43}  +i \Omega_2(z)\rho_{42}\\
    &\quad -i\tilde{\Omega}_2(\mp z)\rho_{23} + i\Omega_1(z)\rho_{41}.  
\end{align}

We then expand the density matrix as a Fourier series. More precisely, we expand the density matrix for the population differences as 

\begin{equation}
    \beta_{ii}\equiv (\rho_{33}-\rho_{ii})=\sum_{n=-\infty}^{\infty}\beta_{ii}^{(n)}e^{inkz},
\end{equation}
and for the coherences we use Eq. \eqref{eq:fourierharmonics}.
The total system consists of 9 equations, but can be further reduced to 7 equations by expressing the coherences $\rho_{13}$, $\rho_{23}$, $\rho_{14}$ and $\rho_{24}$ in terms of the populations and $\rho_{12}$ and $\rho_{34}$. By making this substitution, we arrive at the recurrent matrix equation

\begin{equation}
    A_n\textbf{x}_{n-2}+B_n\textbf{x}_n+C_n\textbf{x}_{n+2}=\mathbf{\boldsymbol{\gamma}} \delta_{n0},  
    \label{eq:recursive}
\end{equation}
defined by the $7\times 7$ matrices $A_n$, $B_n$, $C_n$, and the vectors

\begin{equation}
    \textbf{x}_n=\begin{pmatrix} \beta_{11}^n \\ \beta_{22}^n \\ \beta_{44}^n \\ \rho_{12}^n \\ \rho_{21}^n \\ \rho_{34}^n \\ \rho_{43}^n\end{pmatrix}, \quad \boldsymbol{\gamma}=i\begin{pmatrix} 1 \\ \tfrac{7}{2} \\ -\tfrac{i}{4}\\ 0 \\ 0 \\ 0 \\ 0\end{pmatrix}.
\end{equation}
The explicit forms of $A_n$, $B_n$ and $C_n$ can be found from substitution and are provided in the supplemental data. The continued fractions problem is  solved by repeatedly applying the recursion relation defined in Eq.~\eqref{eq:recursive}.

\begin{figure}
    \centering
    \includegraphics[width=1\linewidth]{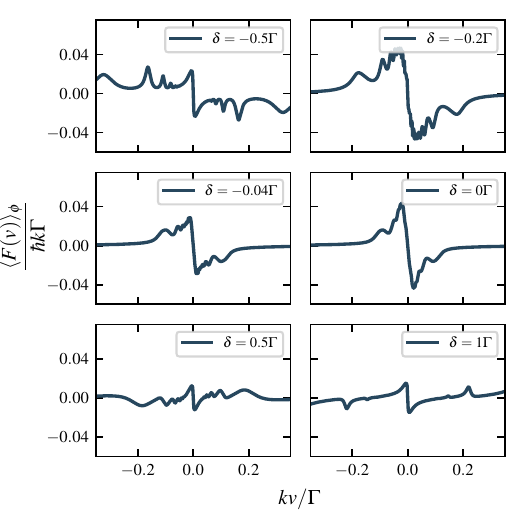}
    \caption{\textbf{Phase-averaged force around the Raman resonance.} The force $\langle F(v)\rangle_{\phi}$ defined in Eq. \eqref{eq:force-phase-avg} is shown for various detunings $\delta$ around the Raman resonance with $\Omega_1=0.58\,\Gamma$, $\Omega_2=1.63\,\Gamma$, $\delta_4=19.9\,\Gamma$ and $\delta_2=5\,\Gamma$. }
    \label{fig:force-velocity2x3}
\end{figure}

In Fig. \ref{fig:force-velocity2x3}, the phase-averaged force defined in Eq.~\eqref{eq:force-phase-avg} is shown for various detunings around the Raman resonance for $\delta_2=5\,\Gamma$. For small and negative detunings $\delta$, a cooling force is observed, which transitions to heating around $\delta=0.5\,\Gamma$. The cooling slope near zero velocity for $\delta=0.5\,\Gamma$ and $\delta=1\,\Gamma$ corresponds to velocities of the order of the recoil velocity, consistent with previous observations \cite{PhysRevA.87.063411}. It should be noted that the heating feature at $\delta=0.5\,\Gamma$ depends on the phase and becomes more pronounced for a smaller ratio $\left(\Omega_1^2/\Omega_2^2\right)<0.1$. Depending on the relative phase, alternating intervals of heating and cooling are observed, which combine in the phase-averaged results.

\newpage
\end{document}